\documentclass[
]{ceurart}

\usepackage{listings}

\usepackage{subfig}

\usepackage{graphicx}
\usepackage{tabularx}
\usepackage{enumitem}
\newboolean{showcomments}
\setboolean{showcomments}{true}
\ifthenelse{\boolean{showcomments}}
 { \newcommand{\mynote}[2]{
      \fbox{\bfseries\sffamily\scriptsize#1}      {\small$\blacktriangleright$\textsf{\emph{#2}}$\blacktriangleleft$}}}
        { \newcommand{\mynote}[2]{}}

\begin{document}

\copyrightyear{2023}
\copyrightclause{Copyright for this paper by its authors.
  Use permitted under Creative Commons License Attribution 4.0
  International (CC BY 4.0).}

\conference{HITLAML'23: Human-in-the-Loop Applied Machine Learning,
  September 04--06, 2023, Belval, Luxembourg}

\title{Can LLMs Demystify Bug Reports?}

%\tnotetext[1]{You can use this document as the template for preparing your publication. We recommend using the latest version of the ceurart style.}


\author[1]{Laura Plein}[%
email=laura.plein@men.lu
]
\address[1]{University of Luxembourg}
\author[1]{Tegawendé F. Bissyandé}[%
email=tegawende.bissyande@uni.lu
]

%\cormark[1]


%\cortext[1]{Corresponding author.}
%\fntext[1]{These authors contributed equally.}

\begin{abstract}
Bugs are notoriously challenging:  they slow down software users and result in time-consuming investigations for developers. These challenges are exacerbated when bugs must be reported in natural language by users. Indeed, we lack reliable tools to automatically address reported bugs (i.e., enabling their analysis, reproduction, and bug fixing). With the recent promises created by LLMs such as ChatGPT for various tasks, including in software engineering, we ask ourselves: What if ChatGPT could understand bug reports and reproduce them? This question will be the main focus of this study. To evaluate whether ChatGPT is capable of catching the semantics of bug reports, we used the popular Defects4J benchmark with its bug reports. Our study has shown that ChatGPT was able to demystify and reproduce 50\% of the reported bugs.
ChatGPT being able to automatically address half of the reported bugs shows promising potential in the direction of applying machine learning to address bugs with only a human-in-the-loop to report the bug.
\end{abstract}

\begin{keywords}
  User Bug Reports \sep
 % Chatbots \sep
  ChatGPT \sep
  LLMs \sep
  Automated Testing
\end{keywords}

\maketitle

\section{Introduction}

Software users are expected to provide feedback on their experience in running programs. 
Such feedback often leads to various improvements by developers responding to feature requests and bug reports. 
In this respect, development platforms, such as GitHub, offer tool support for collecting reports and continuously monitoring how developers address them. 
Unfortunately, various studies have shown that bug reports are under-exploited~\cite{bissyande2013got}. 
Recurrently, indeed, researchers and practitioners point to the general quality of such reports: developers put much effort into ``understanding'' and reproducing the potential bugs that are reported; 
researchers struggle to build  tools for automatically capturing the semantics of the natural language text and transforming them into actionable inputs for existing (testing) frameworks. 

With recent advances in natural language processing techniques, such as the advent of large language models (LLMs), a wide range of tasks have seen machine learning achieve, or even exceed, human performance. Machine translation~\cite{lopez2008statistical,stahlberg2020neural}, in particular, has been a very active field where several case studies have been explored beyond language translation. For example, in software engineering, several research directions have investigated the feasibility of leveraging natural language inputs for producing  programming artefacts and vice-versa. Some milestones have been recorded in the literature in code summarization~\cite{allamanis2018survey,hu2018deep}, program repair~\cite{goues2019automated,monperrus2018automatic}, and even program synthesis~\cite{gulwani2017program}. Nevertheless, bug reports have scarcely been explored. Yet, automating bug reproduction via analysis of bug reports holds tremendous value. In this work, we propose to {\em study the feasibility of exploiting an LLM for reproducing bugs}. We focus on ChatGPT, which has recently received much attention and presents the advantage that its model has been trained on a large corpus of natural language text as well as the source code of software programs. 

But {\em can ChatGPT demystify bug reports?} We consider the management of bug reports as an example case where machine learning can be helpful while keeping the human in the loop. ``Demystifying bug reports'' suggests the eventual possibility of reproducing the reported bug. Our prompt is therefore focused on requesting ChatGPT to exploit a bug report's textual content (in natural language) and generate a formal test case (in a programming language). We assume that if ChatGPT can generate a test case that not only is executable but also fails on the associated buggy program version, then ChatGPT may have ``demystified'' (in the sense of ``captured the semantics of unwanted execution behavior reported by the user'') the content of the bug report. This assumption is, obviously, an over-approximation of the relevance of the generated test case since the generated failing test case may be based on random inputs that are irrelevant to the reported bug. Nevertheless, it would constitute a first milestone towards automatic test case generation based on user inputs, which reflects a realistic and complex user experience.

\section{Related Work}
To address a bug, the first step is to understand it and reproduce it. To demonstrate that there is a bug, the developer has to write a bug-triggering test case since the original test suites are usually scarce and incomplete~\cite{le2019reliability,xiong2018identifying}. Several techniques~\cite{anand2013orchestrated,taneja2008diffgen,thummalapenta2009mseqgen,fraser2013evosuite} have been developed to help developers with this very time-consuming process. However, these techniques mostly rely on formal specifications.

Recently, some work on test case generation using large language models (LLM) was done by~\cite{schafer2023adaptive}, but their TestPilot still requires the functions signature and implementation as prompt. The use of ChatGPT to directly enhance Automated Program Repair (APR) techniques~\cite{xia2023automated,xia2023conversational,sobania2023analysis} highlights even more the potential of this new LLM. Feng et al.~\cite{feng2023prompting} have investigated ChatGPT's ability to help developers reproduce the bug while extracting important steps to reproduce the bug from the bug report. However, they did not perform any test case generation. Additionally, they still require a human to actually reproduce the bug. In contrast, in our study, we are aiming at using the unprocessed human-written bug report as direct input for test case generation, enabling automatic bug reproduction with only a human-in-the-loop to report the bug but not to address it.

\section{Experimental setup}
\label{setup}
\begin{figure}[h!]
\centering
    \includegraphics[width=12cm]{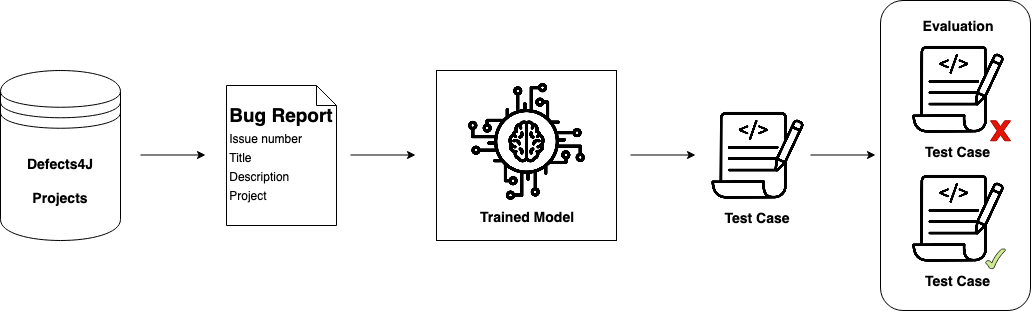}
    \caption{From an informal bug report to a valid test case}
    \label{fig:overview}
\end{figure}

In this section, we overview the settings under which we assess the capability of a Large Language Model to translate informal bug reports from real software projects into formal test case specifications that reproduce the buggy behavior as illustrated in Figure~\ref{fig:overview}. In particular, we present the benchmark, the metrics as well as the experimental design. 

\subsection{Benchmark}
We consider the \textbf{Defects4J} repository~\cite{defects4J-dissection}, which includes real-world faults from various Java software development projects, as enumerated in Table~\ref{table:defects4j}. We collect the bug reports associated to these faults. One must mention that not all bug reports were available, and some bugs referred to the same bug report; in that case, it was only considered once to avoid bias in the results because of duplicates. We considered Defects4J due to its wide adoption in the software testing and software research communities. 

\begin{table}[h!]
\begin{center}
\begin{tabular}{ |c|c|c|c|c| c| c|} \cline{2-7}
\multicolumn{1}{c|}{}  & \multicolumn{6}{c|}{Project}   \\  \cline{2-7}
 \multicolumn{1}{c|}{}  & Chart & Cli & Closure & Lang & Math & Time  \\ 
\hline
 \#Faults & 26 & 39 & 174 & 64 & 106 & 26 \\ 
\hline
 \#Faults associated to Bug Reports & 6 & 30 & 127 & 60 & 100 & 19 \\ 
 \hline
\end{tabular}
\centering
\caption{Java Projects from Defects4J used in this study}
 \label{table:defects4j}
\end{center}
\end{table}

%Note that in Table~\ref{table:defects4j} the amount of bug reports is the amount of different bug reports and therefore not necessarily equal to the amount of bugs of a project. In most projects there were several bugs sharing the same bug report.

\subsection{Metrics}
%\textbf{Test case evaluation} 
As introduced, the goal of the study is to assess whether ChatGPT can generate test cases from bug reports. Therefore, our evaluation is focused on measuring  the quality of the generated test cases. We thus consider two main metrics:
\begin{itemize}
    \item {\bf Executability: } A ChatGPT-generated test case may not even be syntactically correct to be compiled and executed. While often, the generated test case can be made executable after manually implementing small edits (e.g., adding relevant imports), we conservatively consider that executability is a binary metric and is automatically computed once ChatGPT outputs are yielded (with no manual changes added).
    \item {\bf Validity: } An executable test case may or may not fail on the target buggy program. We follow the convention of patch validation in program repair and consider the generated test case to be valid only when it, indeed, fails on the buggy program. Otherwise, it is considered invalid. 
\end{itemize}

\subsection{Experimental Design}
%For this study we consider \textbf{ChatGPT}, which is currently accepted as state-of-the-art for modelling code since it encompasses many knowledge for NLP and code. 
We rely on the ChatGPT API (version 3.5) for our experiments. We construct the prompt by concatenating two pieces of information: the instruction and the bug report. 
The instruction is unique for all queries to ChatGPT and is as follows: ``{write a Java test case for the following bug report: }''. For the bug report, our feasibility study considers that no pre-processing should be applied to the bug report, and the information should not include follow-up comments or attachments. For every prompt, we request ChatGPT five (5) times and assess the different generated test cases. In practice, before running the generated test cases, the ChatGPT outputs are parsed to clean them from natural language texts (e.g., explanations) which would lead to compilation failures. Afterward, the test cases are systematically included in the test suite, which is fully executed by the Defects4J test pipeline. Execution results are then logged, allowing us to compute the metrics on executability and validity.
%directly as baseline to generate test cases.

%This benchmark is well suited for this study since it has been used to evaluate most state-of-the-art APR tools. 
%Another motivation for this dataset is that all included projects (Chart, Closure, Lang, Math, Mockito and Time) contain only Java test cases: having test cases from only one programming language facilitates the model training for test case generation. 

% to fix 
% add accuracy based on similarity with developer written test case? or only for ICSE?

%\indent\textbf{Relevance} is another binary metric which describes whether the generated test case was able to reproduce the bug. To verify the relevance of a test case for a given bug we have the two following criteria. 

%First, the test case must fail on the buggy project version and, additionally, the test case must pass on the fixed project version. Only if both criteria are met, we consider the generated test case as relevant for the given bug report. \\

\section{Results}
\label{results}

\begin{figure}[h!]
\centering
    \subfloat[Bug Report]{
        \includegraphics[width=7cm]{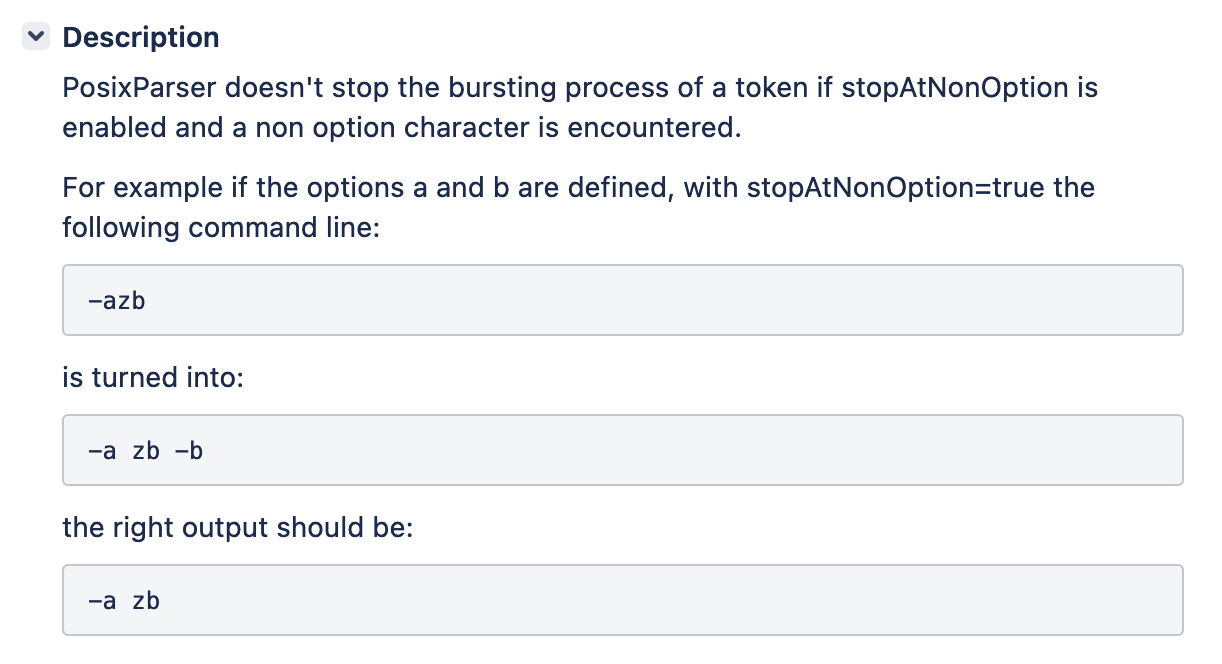}
    }\quad
    \subfloat[Generated Test Case]{%
        \includegraphics[width=7cm]{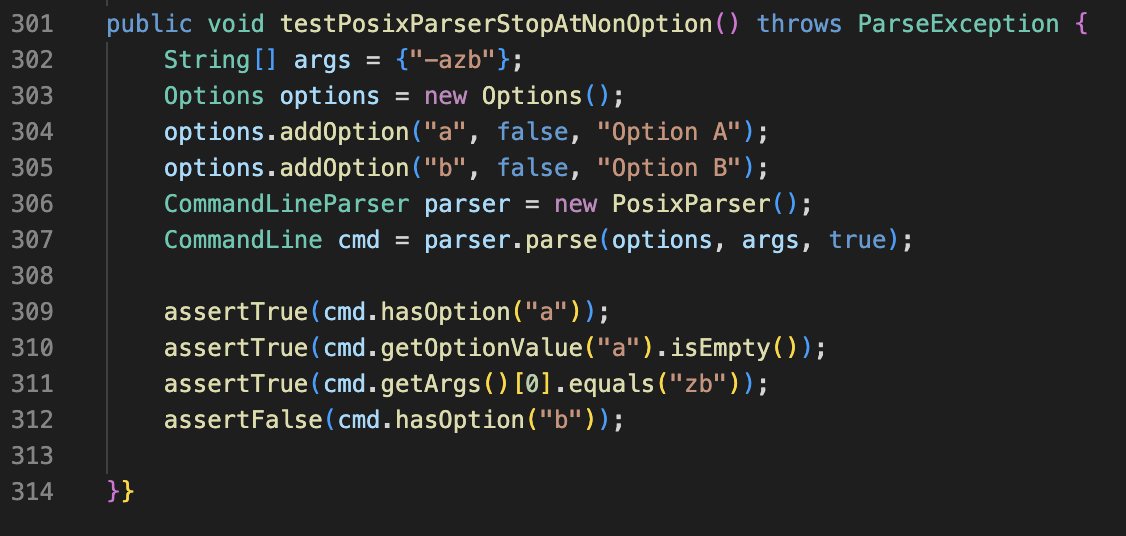}
    }\quad  
    \subfloat[Original Bug-Triggering Test Case]{%
        \includegraphics[width=9cm]{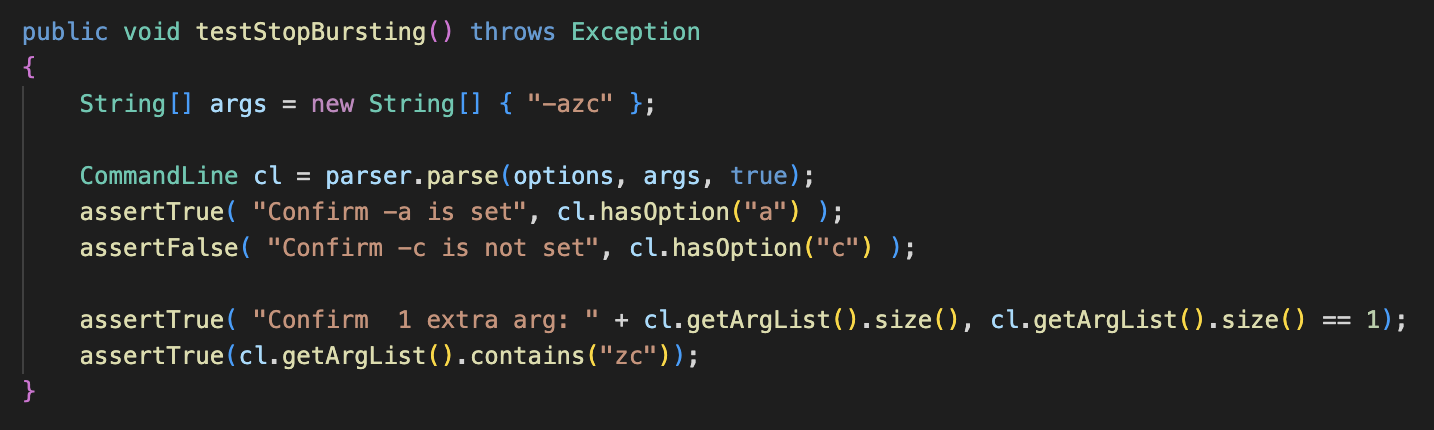}
    }
  \caption{Cli 17 Example from Defects4J}
  \label{fig:gencli}
\end{figure}

Figure~\ref{fig:gencli} provides an illustrative example of a bug report (from the CLI project) and the associated formal test cases (ground truth in Defects4J and generated from ChatGPT). As we can see in this example, ChatGPT is able to reproduce a formal test case from a bug report, which can enable various software automation tasks, such as spectrum-based fault localization, patch validation in program repair, and, more generally, automated software testing. 

%\begin{figure}[h!]
 % \includegraphics[width=4.5cm]{figures/cli_17_bug_report.png}
  %\includegraphics[width=4.5cm]{figures/cli_17_gen_test.png}
  %\includegraphics[width=4.5cm]{figures/cli_17_original.png}
  %\caption{Generated test case for bug report}
  %\label{fig:gencli}
%\end{figure}

On the Defects4J dataset, we compute the proportion of bug reports for which ChatGPT is able to successfully generate test cases.
Table \ref{fig:genPercComb} summarizes the metrics. On average, executable test cases were obtained for 50\% of the bug reports across all projects. The validity of the generated test cases varies greatly from one project to another, which can be explained by the different source and format of user-written bug reports. 
%For some Apache projects, such as Cli, Lang and Math, the bug reports were generally quite informative whereas the GitHub issues of other projects were often just a very short and brief description of the bug. 
Unfortunately, a commonly known issue of some Math project versions is that the test suite is compiling, but their execution is never-ending. Therefore, there were 10 Math bug reports for which we could not determine the \textit{validity} of the generated test cases, which explains the lower validity results for this project. Overall, we got valid test cases for 30\% of the bug reports, which is really promising. Among the executable test cases, we observed that 59\% were valid. This shows that once the initial \textit{executability} challenge is passed, the tests generated by ChatGPT are actually valid, highlighting its understanding. Meaning that ChatGPT was able to catch the semantics of the user-written bug report and translate them into a bug-triggering test case. These results highly motivate further research in this domain.\\

Further manual investigations also highlighted that \textit{executability} and \textit{validity} could, in most cases be fixed with minor modifications (e.g., adding relevant imports or changing duplicated function names). Additionally, some bug reports were very short (only one line) and not descriptive of the problem; those will, in most cases, lead to non-executable and irrelevant test cases.\\ 
% This motivates even further the adoption of our approach. \\  

\begin{table*}[!t]
\begin{center}
\begin{tabularx}{1\textwidth}{  
  | >{\centering\arraybackslash}X 
  || >{\centering\arraybackslash}X
  | |>{\centering\arraybackslash}X 
  | >{\centering\arraybackslash}X 
  | >{\centering\arraybackslash}X | } 
 \cline{3-5}
 \multicolumn{1}{c}{} & \multicolumn{1}{c|}{}  & \multicolumn{3}{c|}{percentage of generation success} \\ 
 \hline
  Project & \# of bug reports  & overall executability & overall validity & validity among executable\\ 
\hline
 Chart & 6  & 33\% & 17\% &50\%\\ 
 Cli & 30 &   53\% & 37\% &69\%\\ 
 Closure & 127  & 46\% & 28\% &59\%\\ 
 Lang & 60 &   60\% & 43\% &72\%\\ 
 Math & 100  & 43\% & 15\% & 35\%\\ 
 Time & 19 &   84\% & 68\% &81\%\\ 
 \hline
 \hline
 Total & 342 &  50\% & 30\% & 59\%\\
 \hline
\end{tabularx}
 \caption{Generation performance for ChatGPT: percentage of bug reports where we successfully generated at least one test case.}
 \label{fig:genPercComb}
\end{center}
\end{table*}

% cli 11 valid (failing on buggy version)
% closure 35 valid
% lang 26 valid
% math 15 valid
% time 13 valid
% total valid: 101

%hope that validity among executable >50% for all...

%\noindent
\fbox{\fbox{\parbox{5.5in}{
\textbf{Findings:} The experimental results based on ChatGPT's API show that a large language model (LLM) can take bug reports as inputs and produce test cases that are executable in 50\% of the cases. Beyond \textit{executability}, about 30\% of the bugs could be reproduced with valid test cases. 
Specifically, over half (59\%) of the executable test cases were valid test cases. 
These results, which are based on an off-the-shelf LLM as-a-service, show promises for automated test case generation beyond unit testing, leveraging complex information from user-reported bugs.}}}

\section{Discussion}
\label{discussion}
Overall, our empirical study validates the hypothesis that ChatGPT can ``\textit{demystify}'' bug reports: given a bug report, it can extract its semantics and translate them into formal test cases. However, some challenges that should be addressed in the future remain. This section will cover some limitations and suggest some research directions for future work to increase the amount of executable and valid test cases.

%\subsection{Related Work}
%To address a bug, the first step is to understand it and reproduce it. To demonstrate that there is a bug, the developer has to write a bug-triggering test case since the original test suites are usually scarce and incomplete~\cite{le2019reliability,xiong2018identifying}. Several techniques~\cite{anand2013orchestrated,taneja2008diffgen,thummalapenta2009mseqgen,fraser2013evosuite} have been developed to help developers with this very time consuming process. However, these techniques mostly rely on formal specifications.

%Recently, some work on test case generation using large language models (LLM) was done by~\cite{schafer2023adaptive}, but their TestPilot still requires the functions signature and implementation as prompt. The use of ChatGPT to directly enhance Automated Program Repair (APR) techniques [~\cite{xia2023automated},~\cite{xia2023conversational},~\cite{sobania2023analysis}] highlights even more the potential of this new LLM. Feng et al.~\cite{feng2023prompting} have investigated ChatGPTs ability to help developers reproduce the bug while extracting important steps to reproduce the bug from the bug report. However they did not perform any test case generation. Additionally, they still require a human to actually reproduce the bug. In contrast, in our study, we are aiming at using the unprocessed human written bug report as direct input for test case generation, enabling automatic bug reproduction with only a human-in-the-loop to report the bug but not to address it.

\subsection{Threats to Validity}
In this study, ChatGPT, an openly available LLM known to have been trained on public data, is exploited to generate test cases for bugs of the Defects4J dataset. As Defects4J is a standard benchmark that is publicly available, it is likely that parts of the dataset have been part of the model's training data; this is a threat to the validity of the results as it reflects a data leakage problem. To address this concern, we performed manual investigations of the generated test cases to ensure their difference from the original tests, confirming ChatGPT's capability of catching the semantics of the bug report. Still, in further iterations of this work, we will assess ChatGPT's performance on newly reported faults. Additionally, due to the randomness of ChatGPT, it is essential to verify that ChatGPT is correctly replying to the given prompt before starting the experiments.

\subsection{Limitations \& Future Work}
In this study, the \textbf{executability} only reflects if a generated test case is directly executable or not. This doesn’t reflect the amount of effort for a human to make it executable. After manually reviewing the generated test cases, it has been shown that most can be made executable through the modification of one or two lines of code. The most common issues are usually: missing imports, duplicate function names, or the use of a deprecated function. Those limitations could systematically be fixed in future work (e.g., with prompt engineering), significantly increasing the number of executable test cases.
For future iterations of this study, we suggest to investigate the potential of \textbf{fine-tuning LLMs} to translate the informal bug reports into formal test cases with a higher rate of executable and valid test cases.
In our experiments, \textbf{bug reports} were collected and directly used as prompts to demonstrate the feasibility and applicability of our idea to address real software faults reported by the users. Nevertheless, pre-processing the textual data might be beneficial to keep ChatGPT focused on the main context of the bug report, therefore increasing the executability and validity of the generated test case.
In order to fix a bug, the first step is to reproduce it with a bug-triggering test case which has been proven feasible in our study. In order to reach to directly fix a newly reported bug, further research should be done on how we can improve and automatically deploy APR tools to not just automatically address the bug but fix it.

\section{Conclusion}
\label{conclusion}
Large language models have recently gained substantial popularity, thanks to the public release of ChatGPT, whose potential as a disruptive technology has been largely advertised. The literature has empirically studied various capabilities of the large language model for various natural language processing tasks. In this work, we investigate the feasibility of leveraging ChatGPT to translate informal bug reports into formal test case specifications. The promising results, in terms of \textit{executability} (i.e., the test case is syntactically correct), and \textit{validity} (i.e., the test case actually makes the program fail), suggest that ChatGPT can ``demystify'' the semantics of bug reports. This finding is essential as it opens new research directions with large language models towards automating test case generation with a human in the loop (for writing bug reports).


%%
%% Define the bibliography file to be used
%\bibliography{sample-ceur}
\newpage
\bibliography{bib.bib}

\begin{thebibliography}{20}
\expandafter\ifx\csname natexlab\endcsname\relax\def\natexlab#1{#1}\fi
\providecommand{\url}[1]{\texttt{#1}}
\providecommand{\href}[2]{#2}
\providecommand{\path}[1]{#1}
\providecommand{\DOIprefix}{doi:}
\providecommand{\ArXivprefix}{arXiv:}
\providecommand{\URLprefix}{URL: }
\providecommand{\Pubmedprefix}{pmid:}
\providecommand{\doi}[1]{\href{http://dx.doi.org/#1}{\path{#1}}}
\providecommand{\Pubmed}[1]{\href{pmid:#1}{\path{#1}}}
\providecommand{\bibinfo}[2]{#2}
\ifx\xfnm\relax \def\xfnm[#1]{\unskip,\space#1}\fi
%Type = Inproceedings
\bibitem[{Bissyand{\'e} et~al.(2013)Bissyand{\'e}, Lo, Jiang, R{\'e}veillere,
  Klein, and Le~Traon}]{bissyande2013got}
\bibinfo{author}{T.~F. Bissyand{\'e}}, \bibinfo{author}{D.~Lo},
  \bibinfo{author}{L.~Jiang}, \bibinfo{author}{L.~R{\'e}veillere},
  \bibinfo{author}{J.~Klein}, \bibinfo{author}{Y.~Le~Traon},
\newblock \bibinfo{title}{Got issues? who cares about it? a large scale
  investigation of issue trackers from github},
\newblock in: \bibinfo{booktitle}{2013 IEEE 24th international symposium on
  software reliability engineering (ISSRE)}, \bibinfo{organization}{IEEE},
  \bibinfo{year}{2013}, pp. \bibinfo{pages}{188--197}.
%Type = Article
\bibitem[{Lopez(2008)}]{lopez2008statistical}
\bibinfo{author}{A.~Lopez},
\newblock \bibinfo{title}{Statistical machine translation},
\newblock \bibinfo{journal}{ACM Computing Surveys (CSUR)} \bibinfo{volume}{40}
  (\bibinfo{year}{2008}) \bibinfo{pages}{1--49}.
%Type = Article
\bibitem[{Stahlberg(2020)}]{stahlberg2020neural}
\bibinfo{author}{F.~Stahlberg},
\newblock \bibinfo{title}{Neural machine translation: A review},
\newblock \bibinfo{journal}{Journal of Artificial Intelligence Research}
  \bibinfo{volume}{69} (\bibinfo{year}{2020}) \bibinfo{pages}{343--418}.
%Type = Article
\bibitem[{Allamanis et~al.(2018)Allamanis, Barr, Devanbu, and
  Sutton}]{allamanis2018survey}
\bibinfo{author}{M.~Allamanis}, \bibinfo{author}{E.~T. Barr},
  \bibinfo{author}{P.~Devanbu}, \bibinfo{author}{C.~Sutton},
\newblock \bibinfo{title}{A survey of machine learning for big code and
  naturalness},
\newblock \bibinfo{journal}{ACM Computing Surveys (CSUR)} \bibinfo{volume}{51}
  (\bibinfo{year}{2018}) \bibinfo{pages}{1--37}.
%Type = Inproceedings
\bibitem[{Hu et~al.(2018)Hu, Li, Xia, Lo, and Jin}]{hu2018deep}
\bibinfo{author}{X.~Hu}, \bibinfo{author}{G.~Li}, \bibinfo{author}{X.~Xia},
  \bibinfo{author}{D.~Lo}, \bibinfo{author}{Z.~Jin},
\newblock \bibinfo{title}{Deep code comment generation},
\newblock in: \bibinfo{booktitle}{Proceedings of the 26th conference on program
  comprehension}, \bibinfo{year}{2018}, pp. \bibinfo{pages}{200--210}.
%Type = Article
\bibitem[{Goues et~al.(2019)Goues, Pradel, and
  Roychoudhury}]{goues2019automated}
\bibinfo{author}{C.~L. Goues}, \bibinfo{author}{M.~Pradel},
  \bibinfo{author}{A.~Roychoudhury},
\newblock \bibinfo{title}{Automated program repair},
\newblock \bibinfo{journal}{Communications of the ACM} \bibinfo{volume}{62}
  (\bibinfo{year}{2019}) \bibinfo{pages}{56--65}.
%Type = Article
\bibitem[{Monperrus(2018)}]{monperrus2018automatic}
\bibinfo{author}{M.~Monperrus},
\newblock \bibinfo{title}{Automatic software repair: A bibliography},
\newblock \bibinfo{journal}{ACM Computing Surveys (CSUR)} \bibinfo{volume}{51}
  (\bibinfo{year}{2018}) \bibinfo{pages}{1--24}.
%Type = Article
\bibitem[{Gulwani et~al.(2017)Gulwani, Polozov, Singh
  et~al.}]{gulwani2017program}
\bibinfo{author}{S.~Gulwani}, \bibinfo{author}{O.~Polozov},
  \bibinfo{author}{R.~Singh}, et~al.,
\newblock \bibinfo{title}{Program synthesis},
\newblock \bibinfo{journal}{Foundations and Trends{\textregistered} in
  Programming Languages} \bibinfo{volume}{4} (\bibinfo{year}{2017})
  \bibinfo{pages}{1--119}.
%Type = Inproceedings
\bibitem[{Le et~al.(2019)Le, Bao, Lo, Xia, Li, and
  Pasareanu}]{le2019reliability}
\bibinfo{author}{X.-B.~D. Le}, \bibinfo{author}{L.~Bao},
  \bibinfo{author}{D.~Lo}, \bibinfo{author}{X.~Xia}, \bibinfo{author}{S.~Li},
  \bibinfo{author}{C.~Pasareanu},
\newblock \bibinfo{title}{On reliability of patch correctness assessment},
\newblock in: \bibinfo{booktitle}{2019 IEEE/ACM 41st International Conference
  on Software Engineering (ICSE)}, \bibinfo{organization}{IEEE},
  \bibinfo{year}{2019}, pp. \bibinfo{pages}{524--535}.
%Type = Inproceedings
\bibitem[{Xiong et~al.(2018)Xiong, Liu, Zeng, Zhang, and
  Huang}]{xiong2018identifying}
\bibinfo{author}{Y.~Xiong}, \bibinfo{author}{X.~Liu},
  \bibinfo{author}{M.~Zeng}, \bibinfo{author}{L.~Zhang},
  \bibinfo{author}{G.~Huang},
\newblock \bibinfo{title}{Identifying patch correctness in test-based program
  repair},
\newblock in: \bibinfo{booktitle}{Proceedings of the 40th international
  conference on software engineering}, \bibinfo{year}{2018}, pp.
  \bibinfo{pages}{789--799}.
%Type = Article
\bibitem[{Anand et~al.(2013)Anand, Burke, Chen, Clark, Cohen, Grieskamp,
  Harman, Harrold, McMinn, Bertolino et~al.}]{anand2013orchestrated}
\bibinfo{author}{S.~Anand}, \bibinfo{author}{E.~K. Burke},
  \bibinfo{author}{T.~Y. Chen}, \bibinfo{author}{J.~Clark},
  \bibinfo{author}{M.~B. Cohen}, \bibinfo{author}{W.~Grieskamp},
  \bibinfo{author}{M.~Harman}, \bibinfo{author}{M.~J. Harrold},
  \bibinfo{author}{P.~McMinn}, \bibinfo{author}{A.~Bertolino}, et~al.,
\newblock \bibinfo{title}{An orchestrated survey of methodologies for automated
  software test case generation},
\newblock \bibinfo{journal}{Journal of Systems and Software}
  \bibinfo{volume}{86} (\bibinfo{year}{2013}) \bibinfo{pages}{1978--2001}.
%Type = Inproceedings
\bibitem[{Taneja and Xie(2008)}]{taneja2008diffgen}
\bibinfo{author}{K.~Taneja}, \bibinfo{author}{T.~Xie},
\newblock \bibinfo{title}{Diffgen: Automated regression unit-test generation},
\newblock in: \bibinfo{booktitle}{2008 23rd IEEE/ACM International Conference
  on Automated Software Engineering}, \bibinfo{organization}{IEEE},
  \bibinfo{year}{2008}, pp. \bibinfo{pages}{407--410}.
%Type = Inproceedings
\bibitem[{Thummalapenta et~al.(2009)Thummalapenta, Xie, Tillmann, De~Halleux,
  and Schulte}]{thummalapenta2009mseqgen}
\bibinfo{author}{S.~Thummalapenta}, \bibinfo{author}{T.~Xie},
  \bibinfo{author}{N.~Tillmann}, \bibinfo{author}{J.~De~Halleux},
  \bibinfo{author}{W.~Schulte},
\newblock \bibinfo{title}{Mseqgen: Object-oriented unit-test generation via
  mining source code},
\newblock in: \bibinfo{booktitle}{Proceedings of the 7th joint meeting of the
  European software engineering conference and the ACM SIGSOFT symposium on The
  foundations of software engineering}, \bibinfo{year}{2009}, pp.
  \bibinfo{pages}{193--202}.
%Type = Inproceedings
\bibitem[{Fraser and Arcuri(2013)}]{fraser2013evosuite}
\bibinfo{author}{G.~Fraser}, \bibinfo{author}{A.~Arcuri},
\newblock \bibinfo{title}{Evosuite: On the challenges of test case generation
  in the real world},
\newblock in: \bibinfo{booktitle}{2013 IEEE sixth international conference on
  software testing, verification and validation}, \bibinfo{organization}{IEEE},
  \bibinfo{year}{2013}, pp. \bibinfo{pages}{362--369}.
%Type = Article
\bibitem[{Sch{\"a}fer et~al.(2023)Sch{\"a}fer, Nadi, Eghbali, and
  Tip}]{schafer2023adaptive}
\bibinfo{author}{M.~Sch{\"a}fer}, \bibinfo{author}{S.~Nadi},
  \bibinfo{author}{A.~Eghbali}, \bibinfo{author}{F.~Tip},
\newblock \bibinfo{title}{Adaptive test generation using a large language
  model},
\newblock \bibinfo{journal}{arXiv preprint arXiv:2302.06527}
  (\bibinfo{year}{2023}).
%Type = Inproceedings
\bibitem[{Xia et~al.(2023)Xia, Wei, and Zhang}]{xia2023automated}
\bibinfo{author}{C.~S. Xia}, \bibinfo{author}{Y.~Wei},
  \bibinfo{author}{L.~Zhang},
\newblock \bibinfo{title}{Automated program repair in the era of large
  pre-trained language models},
\newblock in: \bibinfo{booktitle}{Proceedings of the 45th International
  Conference on Software Engineering (ICSE 2023). Association for Computing
  Machinery}, \bibinfo{year}{2023}.
%Type = Article
\bibitem[{Xia and Zhang(2023)}]{xia2023conversational}
\bibinfo{author}{C.~S. Xia}, \bibinfo{author}{L.~Zhang},
\newblock \bibinfo{title}{Conversational automated program repair},
\newblock \bibinfo{journal}{arXiv preprint arXiv:2301.13246}
  (\bibinfo{year}{2023}).
%Type = Article
\bibitem[{Sobania et~al.(2023)Sobania, Briesch, Hanna, and
  Petke}]{sobania2023analysis}
\bibinfo{author}{D.~Sobania}, \bibinfo{author}{M.~Briesch},
  \bibinfo{author}{C.~Hanna}, \bibinfo{author}{J.~Petke},
\newblock \bibinfo{title}{An analysis of the automatic bug fixing performance
  of chatgpt},
\newblock \bibinfo{journal}{arXiv preprint arXiv:2301.08653}
  (\bibinfo{year}{2023}).
%Type = Article
\bibitem[{Feng and Chen(2023)}]{feng2023prompting}
\bibinfo{author}{S.~Feng}, \bibinfo{author}{C.~Chen},
\newblock \bibinfo{title}{Prompting is all your need: Automated android bug
  replay with large language models},
\newblock \bibinfo{journal}{arXiv preprint arXiv:2306.01987}
  (\bibinfo{year}{2023}).
%Type = Inproceedings
\bibitem[{Sobreira et~al.(2018)Sobreira, Durieux, Madeiral, Monperrus, and
  Maia}]{defects4J-dissection}
\bibinfo{author}{V.~Sobreira}, \bibinfo{author}{T.~Durieux},
  \bibinfo{author}{F.~Madeiral}, \bibinfo{author}{M.~Monperrus},
  \bibinfo{author}{M.~A. Maia},
\newblock \bibinfo{title}{Dissection of a bug dataset: Anatomy of 395 patches
  from defects4j},
\newblock in: \bibinfo{booktitle}{Proceedings of SANER}, \bibinfo{year}{2018}.

\end{thebibliography}


\end{document}